\documentstyle[twocolumn,aps,eqsecnum]{revtex}

\begin{document}
\draft

\title{Chiral patterns arising from electrostatic growth models}

\author{Ilya M.\ Sandler$^{1,2}$, Geoffrey S.\ Canright$^{1,2}$, 
Hongjun Gao$^{3,2}$, Shijin Pang$^3$, Zengquan Xue$^4$,
and Zhenyu Zhang$^{2,1}$}
\address{$^1$Department of Physics and Astronomy, University of Tennessee,\\
Knoxville, Tennessee 37996-1200, USA \\
$^2$Solid State Division, Oak Ridge National Laboratory, \\
Oak Ridge, Tennessee 37831-6032, USA \\
$^3$Beijing Laboratory of Vacuum Physics, Chinese
Academy of Sciences, \\
Beijing 100080, PRC \\
$^4$Department of Electronics, Peking University, Beijing 100871,
China}

\maketitle

\begin{abstract}

Recently, unusual and strikingly beautiful seahorse-like growth patterns
have been observed under conditions of quasi-two-dimensional growth. 
These `S'-shaped patterns strongly break two-dimensional
inversion symmetry; however such broken symmetry occurs only at the 
level of overall morphology, as the clusters are formed from achiral
molecules with an achiral unit cell. 
Here we describe a mechanism which gives rise to chiral growth
morphologies without invoking microscopic chirality. 
This mechanism involves trapped electrostatic charge on the growing
cluster, and the enhancement of growth in regions of large electric
field. We illustrate the
mechanism with a tree growth model, with a continuum model for the
motion of the one-dimensional boundary, and with microscopic Monte Carlo
simulations. Our most dramatic results are found using the continuum
model, which strongly exhibits spontaneous chiral symmetry breaking,
and in particular finned `S' shapes like those seen in the experiments.

\end{abstract}


\section{Introduction}

Growth phenomena are known for the complexity and beauty of the patterns they
can lead to \cite{fractalGrowth}. Most of this complexity results from
different kinds of instabilities associated with growth,   such as
the Mullins-Sekerka  instability of growth fronts \cite{mullins}, or fingering
instability \cite{fingering0,fingering1,fingering2}.
The presence of instabilities implies that a tiny microscopic noise can result
in macroscopic changes of shape, and hence lead to a variety of shapes.
(For instance, the  formation of snowflakes, a growth phenomenon familiar to 
everyone, produces hundreds of different shapes \cite{snowFlakes}). 
However, despite the variety of shapes, the vast majority of growth patterns 
preserve left-right symmetry; in other words, essentially all of
the known growth patterns are {\it achiral}. 
One well-known and historic exception is the appearance of hemihedral
faces on crystals, yielding facetted forms which are not invariant under
inversion \cite{JCW}. In the mid-19th century, such faces were identified
in molecular crystals of sodium ammonium tartrate by Pasteur, and the
broken chiral symmetry was ascribed by him to the 
microscopic chirality of the constituent molecules.

In this work we concentrate on two-dimensional (or quasi-two-dimensional)
growth forms. For two-dimensional forms the relevant inversion operator
is $x \to -x$ {\it or} $y \to -y$, but not both; we will call such an
operation ``2D inversion", and forms distinguishable from their 2D
inverse ``2D chiral". (Also, since we concentrate entirely on 2D
henceforth, we will sometimes shorten these terms by omitting the leading
``2D" qualifier.) 2D chiral growth forms are not common \cite{uncommon}.
In those rare cases where
chiral growth patterns do appear  \cite{phospholipid,bacteria}, the inversion
symmetry is already broken at the microscopic level.
An example is the formation
of spiral crystals \cite{phospholipid} 
during the compression of a phospholipid monolayer on
a water-air interface. In this case the individual phospholipid molecules
possess a 3D chirality. Each molecule also has a preferential orientation
(hydrophilic head down) with respect to the water-air interface. 
This consistent orientation then gives a monolayer with 
{\it two-dimensional\/} inversion
symmetry already broken \cite{3Dto2Dnote}
at the microscopic level---assuming only that
the molecules themselves have a predominance of a single enantiomer.
And in fact,
for spiral crystals to appear, one needs to have a monolayer consisting
predominantly of a single enantiomer. The handedness of the crystals depends
directly on the handedness of the dominant enantiomer, and no chirality
appears for racemic monolayers \cite{phospholipid}. 
While there are several competing explanations of how the microscopic
chirality leads to the macroscopic chirality 
\cite{phospholipid1,phospholipid2}, it is nevertheless clear that
the latter occurs only because of the former. Similarly, in another chiral
growth example---the formation of chiral bacterial colonies 
\cite{bacteria}---the individual
particles (bacteria) also have a three-dimensional chirality (of a single
`sign') which  then manifests itself as a 2D chirality when coupled with
a 2D substrate. The bacterial aggregates are observed to be (2D) chiral,
and always with the same handedness \cite{bacteria}.
Thus, in each of these cases, it is clear that the macroscopic
2D chirality of the aggregates
results from a microscopic 3D chirality of the elementary building
blocks.

Recently, a novel and very beautiful type of growth pattern has 
been reported \cite{gao1,gao2}.
A typical pattern strikingly resembles a seahorse (in the form of an `S'
shape, with `fins' on the outer curved edges), and so has a strongly broken
2D inversion symmetry. The patterns were discovered during 
growth studies of fullerene-tetracyanoquinodimethane (C$_{60}$-TCNQ) thin films.
Subsequently, very similar
patterns were obtained using TCNQ only \cite{gao3}.
The broken symmetry is one of the most striking aspects of the patterns,
as well as one of the principal mysteries connected with them.
The mystery arises because---in contrast to 
the two cases mentioned above---in 
these experiments there is {\it no\/} microscopic
symmetry breaking: TCNQ molecules are themselves inversion symmetric
\cite{TCNQstructure}. Furthermore, 
even though the `seahorse' aggregates are polycrystalline \cite{gao1},
one can probably rule out symmetry breaking at the level of the unit cell, since
TCNQ crystals are also achiral \cite{TCNQstructure,probably}. 

It is however important to note that both left-
and right- handed patterns appear in approximately equal numbers
\cite{gao3}. Thus, on average, the experiment does not break 
inversion symmetry; instead 
the symmetry is broken {\it spontaneously}, for each island,
during the growth. That is, the ``seahorse'' growth experiments represent
an almost unique case of spontaneous 2D chiral symmetry breaking during
quasi-two-dimensional growth.

We say `almost' unique because we are aware of only one other growth phenomenon
exhibiting such spontaneous symmetry breaking, namely, phyllotaxis:
the pattern of leaves, buds, scales, etc in growing plants
\cite{phyllotaxis}. As demonstrated by Douady and Couder \cite{Douady},
this phenomenon can be understood cleanly in two dimensions;
and furthermore, the resulting spiral growth patterns are clearly
chiral, and the symmetry breaking is clearly spontaneous
\cite{phyllotaxis,Douady}. Outside of this one example from 
botany, however, we know of
no example of two-dimensional growth---experimental or theoretical---in 
which the resulting growth
patterns spontaneously break 2D inversion symmetry.

In this work we construct a growth model which does yield such spontaneous
symmetry breaking.
More precisely, we consider
a set of models, all embodying the
same ideas. These ideas involve a novel form of long-ranged branch competition
and growth, arising from electrostatic effects. 
We have found that such a mechanism can lead to growth forms
which spontaneously break two-dimensional inversion symmetry.

The growth models which we will consider share several important properties.
These properties are simple, and can be formulated independently
of the nature of the underlying physical processes. 
The physical picture which we consider involves the following elements:
(i) branching, that is, every growing branch should eventually
give rise to new branches; (ii) strong branch competition---in fact, the 
competition has to be so strong, that only 2 main branches ``survive''; 
(iii) long range branch repulsion:
the two branches need to ``feel'' one another and curve away from each other.

Branching is a very common property in growth phenomena \cite{fractalGrowth}.
Branch competition is also very common. It is usually caused by  
screening---that is, by the competition between growing branches 
for incoming particles. But competition due to screening alone
is not strong enough to lead to two-armed shapes. For instance, the 
diffusion-limited aggregation (DLA) model leads to clusters having 4 or
more branches \cite{WS,Family,DLAbranching0,DLAbranching1,DLAbranching2}.
Competition for incoming particles may also cause some branch repulsion. But
this effect is obviously very short-ranged (a branch ``feels'' only its 
neighbors).  Thus in order to achieve (i)--(iii), one needs to introduce 
a long-ranged interaction into the system.  As will be discussed 
later in section \ref{sec:continuum},
electrostatic forces may play an important role in 
the formation of the seahorse patterns. Hence,
in  our models, the long range interaction between branches is also of
electrostatic origin.

We will start with a very simple deterministic
``tree'' growth model which has properties
(i)--(iii) by construction, and show that this model prefers chiral rather
than symmetric shapes. We will then consider a more realistic
quasi-equilibrium continuum model, in which  properties (ii)--(iii) 
arise naturally due to electrostatic interactions. This model 
yields two-armed, finned, S-shaped forms for a range of growth parameters,
for essentially the same physical reasons as does the tree model.
We also report some preliminary studies involving the same physical
ideas but using a microscopic Monte-Carlo approach.
These modeling efforts are inspired by the puzzling and remarkable
experimental patterns; and they yield qualitatively similar growth
forms. It is also encouraging that further growth experiments
involving a static in-plane electric field
(which we discuss briefly below, and in detail in another paper
\cite{gao3}) have provided support for our ideas.

\section{Charged Trees}

In this section we demonstrate, using a simple and highly schematic model,
how long-ranged branch repulsion and 
competition may cause chiral symmetry breaking.
We will formulate a simple growth
model where such branch repulsion and competition are 
present by construction, and show that chiral `S' shapes are preferred 
energetically over symmetric shapes.

A charged tree model is constructed as follows (Fig.~\ref{fig:treegrowth}). 
A tree starts as
as a single charged rod. The ends of this rod are considered to be ``alive''.
Then each alive branch emits
two branches: one branch of length $l_0$ and the other of length $l_1$.
Both new branches grow at a predefined angle $\theta$. All three 
quantities---$l_0$, $l_1$, and $\theta$---are the same for opposite ends,
and do not vary during the growth. When $l_0 \neq l_1$,  there
are 4 possible combinations of growth on every step; among these,
the model chooses
the tree with the lowest electrostatic energy. Then the longer
of the newly added branches become new ``alive'' branches,  
the shorter ones ``die'',
and the process is repeated again. If two (or more) configurations have
the same energy then the selection is done randomly. 
The first 2 steps of growth are shown in Fig.~\ref{fig:treegrowth}.

To complete this model we need to specify how we will compute the electrostatic
energy, as the energy of a 1D charged rod diverges. The most obvious way to
deal with this problem is to
assign some small (but finite) width $w$ to the branches, with
this width satisfying  $w << l_0, l_1$. This allows us to work with
2D, rather than 1D, charge density. To compute the electrostatic energy of
the tree, we further break  the branches  into smaller
(linear) pieces. The $i$th piece has the  length $l_i$, and the 
linear charge density $\lambda_i$, which
is in an obvious way related to 2D charge density $\sigma_i=\lambda_i /w$.

We will consider 2 possible charge distributions:  
1) a conducting charge distribution
$U_i=const$, where $U_i$ is the potential at the center of the $i$-th piece; 
2) a uniform charge distribution $\sigma_i=const$.
In both cases the total charge of the system will be normalized by
the requirement that the average linear charge density be equal to unity,
that is
\begin{equation}
  \bar{\lambda}=\sum\limits_i \lambda_i l_i/ L =1 \label{eq:normalizationForLambda}
\end{equation}
where $L=\sum\limits_i l_i$ is the total length of the structure.

To find the charge densities for each piece we 
construct a set of linear algebraic equations

\begin{equation}
    \sum \limits_{j=0}^{N-1}U_{ij}\lambda_j= 1 \label{eq:systemForLambda}
\end{equation}
where $i=0,\ldots,N-1$,
and $U_{ij}$ is the electrostatic potential which would be
induced at the center of the $i$-th piece by the $j$-th piece if
the latter had a {\it unit} charge density. Clearly the solution 
$\{\lambda_i\}$
of Eq.~(\ref{eq:systemForLambda}) meets the requirement $U_i=const$ 
(the constant was set to  1); this solution 
can then be easily rescaled to meet the normalization 
condition (\ref{eq:normalizationForLambda}).

Using elementary electrostatics, one can show that
the constants $U_{ij}$ are given by 
\begin{equation}
U_{ii} =2   \ln (5.44 w/l_i)
    \label{eq:selfUforPiece}
\end{equation}
and for $i \neq j$
\begin{equation}
U_{ij}=  \ln {\sqrt{(l_j/2-x_i)^2+y_i^2} +(l_j/2-x_i) \over
\sqrt{(l_j/2+x_i)^2+y_i^2} -  (l_j/2+x_i)}          \label{eq:uOfAPiece}
\end{equation}
where $l_j$ is the length of $j$-th piece, $x_i, y_i$ are the coordinates
of the center of the $i$-th piece (the origin is assumed to be at the center of
the $j$-th piece, and the $x$-axis directed along the $j$-th piece).

The growth of such trees was studied numerically, and
all trees grown according to these rules demonstrate some chirality.
One of the typical S-like shapes is shown in Fig.~\ref{fig:besttree}.

Why does the tree prefer to break the left/right symmetry? 
To answer this question, we note that, by introducing 
the repulsive electrostatic interaction, we
effectively made the tree keep its branches as far away from each other 
as possible.
This observation alone accounts for the fact that the tree ``chose'' the 
`S' shape on the first split. However, during  subsequent splits the result
is determined by the interaction between {\it all} branches, and the outcome
depends crucially on how the charge is distributed over the tree: the more
charge is concentrated on the small ``dead'' branches, the stronger is the
symmetry breaking (i.e. the higher is the achievable curvature 
of the main arms).
We can illustrate this point by considering the case of a
uniform charge distribution
$\lambda_i=const$. In this case the potential energy of a tree is given by
\begin{equation}
   U_u={1 \over 2} \sum\limits_{i,j} U_{ij}\lambda_j\lambda_i l_i =
{\lambda ^2 \over 2}\sum\limits_{i,j} U_{ij} l_i
\end{equation}
We have grown a number of trees using the above rules, but with the
assumption of a uniform charge distribution.
The resulting trees have a much more weakly broken symmetry,
which comes primarily from the symmetry breaking at the first
branching; and they are not S-shaped.

This result is consistent with the idea that a higher charge
density on the external ``dead'' branches (which results for instance from 
the conducting charge distribution) enhances the overall chirality of 
the cluster.
We will return to the question of how the charge redistribution influences
which branches survive or die in the next section.

We also note that one can construct a non-deterministic charged tree growth
model, where the growth rates themselves are determined by
the electrostatic interaction between branches. This non-deterministic
growth model also leads to a spontaneous chiral symmetry breaking 
\cite{canright}.

\section{The Continuum Model}

\subsection{Construction of the Model}
\label{sec:continuum}
In this section we will consider a more realistic growth model, in which
strong branch competition and  survival of only 2 main arms occur naturally 
(for some range of parameters). That is, for the model we now 
describe, properties (ii)--(iii) 
(strong branch competition and repulsion)
arise as a  result of the {\it dynamics\/} of growth.
In common with the tree models described above, a crucial
ingredient is the presence of a long-ranged electrostatic interaction.

First we will consider in more detail what actually happens in the ``seahorse''
experiments \cite{gao1,gao2}. In these experiments  layers of TCNQ 
are deposited using the ionized cluster beam (ICB) 
deposition \cite{ICB1,ICB2} method. With this method
the TCNQ molecules are ionized and then accelerated towards 
the substrate,
where they arrive with high kinetic energy (and thus
high mobility) and therefore can diffuse along the substrate and form growing
clusters. A small fraction $\alpha$ of the diffusing particles 
(estimated \cite{gao3}
to be typically $\sim 10\%$) are charged (all of the same sign). 
Thus the growing islands will also carry some (time-dependent) charge. 
The magnitude and time dependence of this charge are not known;
it depends on many complicating factors, including the repulsion of the
charged, diffusing particles by the charge on the island; leakage to the 
substrate, both from the diffusing charged walkers and from the charged
aggregate; and the ``rain'' of charged particles directly on the growing
islands. We will treat this time-dependent charge $Q(t)$ in an extremely
simple way below; our motivation is to explore the kinds of effects that
electrostatic charge may have on growth processes.

The field of the island, in contrast with the field of walkers,
is not random, and therefore will play the dominant role
in how the particles (walkers) diffuse. 
Hence in our work we neglect the random field of the walkers. 
We assume that neutral walkers have a non-zero
polarizability; hence the diffusive motion of both charged and neutral
walkers is affected by the electrostatic field of the aggregates.
To consider the simplest case, 
we will neglect any possible cluster-cluster interactions and consider
only the growth of an isolated island.

An obvious approach to this problem would be
a Monte Carlo (MC) simulation. However, a simple estimate
of the number of particles in seahorse clusters in the
experiments  gives $10^6$--$10^7$ particles.
A direct microscopic 
Monte-Carlo simulation for a problem of this size is very hard,
if not impossible. We have performed some limited MC studies (involving
much smaller particle numbers, ie $N\sim 10^3$), which we
will describe briefly in section \ref{sec:MC}.
Here we will consider an alternative approach, in which, instead of tracing 
the motion of
individual particles, we will compute local growth rates
for the cluster boundary, which we treat as a continuous 1D curve. 
We will obtain equations for the motion of this 1D curve, and study the
kinds of growth that result.

First, let us consider a growing island surrounded by diffusing
walkers. If a walker hits the island, then with some probability
$p_s$ (``sticking'' probability) it (the walker) becomes a part of the island.
Then the local growth rate is 
\begin{equation}
\vec G(x,y,t)={d\vec h(x,y) \over dt}=  \hat n(x,y) \delta_m^2{d^2 N_{hits}(x,y) \over dt\, dl}
p_s 
\end{equation}
where $d\vec h(x,y)$ is the displacement of the given boundary
point $(x,y)$ during the time $dt$, $\hat{n}(x,y)$ is
the unit vector normal to the boundary, $\delta_m$ is a typical intermolecular
distance in the growing cluster (thus, $\delta_m^2$ is the area occupied by
a single molecule), and $ d^2 N_{hits}(x,y)/( dt\, dl) $ is the number of 
hits per unit boundary length per unit time. We then take
\begin{equation}
{d^2 N_{hits}(x,y) \over dt\,dl} = \alpha_{hits} N(x,y) v_T
\end{equation}
where $N(x,y)$ is the walker concentration
near the point $(x,y)$ on the boundary, 
$v_T$ is an average thermal velocity of the walkers,
and $\alpha_{hits}$ is a numerical factor of order of unity. Thus, we can
rewrite the equation for local growth rates as
\begin{equation}
\vec G(x,y,t)={d\vec h(x,y) \over dt}=  \hat n(x,y) G_T p_s  N(x,y) 
\end{equation}
where $G_T=\alpha_{hits} v_T \delta_m^2$ is a constant which depends 
only on the temperature.
\def\N0{N^{(0)}}

Here we will assume that the growth is slow enough to be considered as a 
quasi-equilibrium process (which is the case when $p_s \ll 1$). Our
motivations for this assumption are twofold: 
first, it is physically motivated, in that the sticking
probability may indeed by very small, due to the high kinetic energy
of the walkers in an ICB experiment;
and second, it makes the problem 
tractable, giving a simple form for $N(x,y)$ which enables us to 
concentrate on the motion of the boundary.
Given this assumption, then,
the concentration of walkers is given by 
a quasi-equilibrium Boltzmann distribution:
$ N(x,y,t) = \N0 \exp[-U(x,y,t)/kT]$
where $U(x,y)$ is the potential energy of a walker at the point $(x,y)$, $\N0$
is the concentration far away from the cluster, and
$k$ is Boltzmann's constant.

In our case there are 2 different kinds of particles present in the system:
charged and neutral walkers, each kind having a different concentration and a 
different potential
energy function. The overall growth rate is given by
\begin{equation}
dh(x,y,t)/dt = dh_n/dt + dh_c/dt         \label{eq:totalGrowthRate}
\end{equation}
where the subscript $n$ denotes neutral walkers, and the 
subscript $c$ charged walkers.
The walker concentrations are given by
\begin{equation}
 N_\beta(x,y,t) = \N0_\beta \exp[-U_\beta(x,y,t)/kT] 
\end{equation}
with $\beta=n$ or $c$ and (as discussed above) $\N0_c \ll \N0_n$.
The potential energy $U(x,y)$ is equal to $-\chi E^2/2$ for neutral walkers
(of polarizability $\chi$) in a field $E = |\vec{E}(x,y,t)|$,
and $V(x,y,t)e$ for charged walkers.

Now we assume that the cluster is conducting. As with the
quasiequilibrium assumption, our reasons are both computational and
physical: the conducting cluster is rather straightforward to treat
numerically (and even analytically in some special cases); 
but also, from our charged-tree studies, we expect that 
a conducting cluster will enhance the type of branch competition 
which we wish to study here. For a conducting cluster,
$U_c$ (and hence $dh_c/dt$) are each independent of position on the island.
The applicability of this assumption to the seahorse experiments will be
discussed below.

The model we have constructed thus far requires the computation of the
electric field due to the charge on a two-dimensional
growing cluster, which is in general of an irregular shape.
This electric field is determined by the charge distribution $\sigma(x,y,t)$
on the island. However, $\vec{E}(x,y)$ diverges near the edge of any 2D 
charge distribution; hence, instead of using the field at the edge, we will
use the field at a small (molecular) distance from the edge. 
That is, instead of
$\vec{E}(x,y,t)=\vec{E}(\vec{r},t)$ we will use 
$\vec{E}(\vec{r}+\delta\cdot\hat{n}(x,y),t)$, where $\delta$ is the 
`sticking distance'---the distance
at which a diffusing particle sticks to the
cluster; we will assume $\delta \sim \delta_m$.
The electric field is determined by the shape of the island,
to within an overall scale factor given by $Q$,
the net charge on the island. 

We can now introduce a simpler charge non-conserving model.
The majority of particles are neutral; furthermore, the charged walkers tend to
be repelled from the charged cluster, while the neutral, polarizable
walkers are drawn to it. Hence one can expect that most of
the growth will result from the aggregation of neutral particles.
Therefore, we neglect the term due
to the charged particles in Eq.~(\ref{eq:totalGrowthRate}). The charge effects 
are taken into account simply by rescaling $Q(t)$.
As discussed above, the likely behavior of $Q(t)$ is unknown, and
dependent upon many competing effects. Here
we will use the simplest possible rescaling rule: we will assume that
$Q(t)=A(t) \overline{\sigma}$, where $A(t)$ the overall area of the cluster
and the average charge density  $\overline{\sigma}$ is assumed to be constant.

We now have a sufficient set of ingredients for a complete growth model.
That is, a given initial shape determines the charge
distribution and hence the electric field. The latter then allows
a growth increment in time $dt$ to be computed, yielding a new shape.
With the new shape one then computes a new charge distribution and
field, and so on.

\subsection{Analysis and Simulation}

Now we will show that, with the model as stated, 
a growing island will eventually transform
from a compact to an elongated shape. Consider an island of elliptical
shape, with principal axes $a$, $b$.
If we compute the new boundary, using Eq.~(\ref{eq:totalGrowthRate}), the
new shape will not be exactly elliptical; however, the ellipse's
field may still be used as an approximation. It is then obvious
that this new ellipse will have greater eccentricity than
the original one if
\begin{equation}
{a+  G(a,0)dt\over b +  G(0,b) dt}>{a \over b} \ \ ,
\end{equation}
which (considering only growth from neutral walkers) is equivalent to
\begin{equation}
{G(a,0) \over G(0,b)}=\exp\biggl\{ {\chi \over 2}\Bigl[ E^2(0)-E^2(\pi/2)\Bigr]
/kT\biggr\} > a/b
\ \ . \label{eq:2armInstability}
\end{equation}
The  electrostatic field $E(\phi)$ at distance $\delta$ from
the point $(a\cos \phi, b\sin \phi)$ is given by 
\begin{equation}
 E^2(\phi)= {Q^2\over 2 \delta ab} (b^2\cos^2\phi+a^2 \sin^2\phi)^{-1/2} 
\end{equation}
After introducing the average charge density
$\bar \sigma=Q/A$, where $A=\pi ab$ is the total area of
the cluster, Eq.~(\ref{eq:2armInstability}) becomes
\begin{equation}
\exp\Bigl[ {\chi \bar \sigma^2 \pi^2\over 2 kT \delta}
(a-b)\Bigr] > {a \over b}   \label{eq:2armInstability1}
\end{equation}

It is obvious from this equation that, if at some time during
the growth the relationship (\ref{eq:2armInstability1}) becomes
valid, it will remain valid afterwards.
Initially, the shape is compact (circular), i.e. $a=b=R$. However, due
to the stochastic aspect of the
growth there are always small variations in the radius.
One can expect that the magnitude of these variations is
at least $\sim \delta$, that is $|a - b| \sim \delta$.
The transition from the compact to elongated shape occurs
when the compact shape becomes unstable with respect to such
variations. This happens  when $R$ reaches some critical value $R_0$,
determined by
\begin{equation}
1+\delta /R_0 = \exp({\chi \bar \sigma^2 \pi^2\over 2 kT })
\end{equation}
or
\begin{equation}
R_0 \approx 2 \delta   kT /\chi \bar \sigma^2\pi^2 \ \ .
\end{equation}

Now, estimating $\bar \sigma \sim \alpha e/\delta_m^2$,
$\chi \sim 2\times 10^{-24} {\rm cm}^3$,
$\delta_m \approx \delta\sim 10^{-7}$ cm,
$\alpha=10\%$, and $T \approx 300K$, we get the estimate
$R_0 \sim 0.2 \mu$m. Although this estimate is very rough, it is
encouraging to see that $R_0$ is smaller (by roughly an order of
magnitude) than the size of seahorses observed in the experiments
\cite{gao1,gao2}.

Thus we find an instability of a compact, circular cluster to an 
elliptical form, when the compact cluster exceeds a 
critical size. We have studied the growth of ellipses numerically, and
verified that the elongation decays for $R<R_0$; while for $R>R_0$
the elongation persists and grows well beyond the linearized form,
and in fact is amplified (and `pinched') by the resulting growth, to 
give two arms. We have also numerically tested instabilities to four
arms. Here we find, for the parameters that we have explored,
that the two-arm instability is dominant over four-arm
instabilities.

It is clear that, for most cases of interest, neither the charge density
nor the electric field can be computed analytically. Hence we need
to implement our growth model numerically. This can be done as follows.
The current boundary of the cluster is represented as
a set of vertices $\{\vec r_v:v=0\ldots N_b-1 \}$, connected by
line segments. For each
vertex we compute the potential energy of a walker near this
point $ U(\vec r_v)$; then a unit normal vector $\hat n_v$, and
the displacement vector
$\vec \Delta_v=\hat n_v G(\vec r_v) \Delta_t $
are computed
(where $\Delta_t$ is a time-scaling constant). After all displacements are
computed, all vertices are moved to their new positions.
The only remaining problem is to compute the field distribution
near the boundary.

One can try to deal with this problem by solving straightforwardly
Laplace's equation for the electrostatic potential. However, despite the
fact that our problem is 2-dimensional, Laplace's equation
for the electrostatic potential is still 3-dimensional. Hence,  to find
the local electrostatic field, one would  need to solve this 3D problem
for the full space. To take
advantage of the lower dimension of the problem, we will use a surface
charge method \cite{SCM1,SCM2,SCM3,SCM4}, which we will outline below.

Given the cluster $C$, we first compute the 2D charge density
$\sigma(x,y)$, and then find the electrostatic field.
To find $\sigma(x,y)$, we introduce a grid
$\{\vec D_i,i=1,\ldots,N_g;\vec D_i\in C\}$, forming a square lattice inside the
cluster. Each square of this lattice is assigned a charge
density $\sigma_i$ (we will refer to these squares as elements).
Now we replace the continuous equation for
the electrostatic potential $V(x,y)={\rm constant}$ for all $(x,y) \in C$
by the discrete equation $V(\vec D_i)={\rm constant}$, which gives rise
to the following set of linear algebraic equations:
\begin{equation}
    \sum \limits_{j=0}^{N_g-1}U_{ij}\sigma_j= 1 \label{eq:CMsystemForSigma}
\end{equation}
where $i=0,\ldots,N_g-1$, and the scaling constant on the right-hand
side (here set to 1)
can be chosen arbitrarily (that is, after the system is solved,
we can rescale all the $\sigma_i$).
$U_{ij}$ is the electrostatic potential which would be
induced at the point $\vec D_i$ by the square
element $j$, if the latter had a {\it unit} charge density.
Therefore, one can approximate $U_{ij}$ as follows
\begin{equation}
U_{ij}={A_j \over r_{ij}},\ {\rm for}\ i \neq j \label{eq:pointChargeUij}
\end{equation}
where $A_j$ is the area of the $j$-th element, 
and $r_{ij}=|\vec D_j - \vec D_i|$.
For most purposes the accuracy of this point-charge
equation is sufficient, but one can also use more accurate expression for
$U_{ij}$ which takes into account a quadrupole correction
\begin{equation}
U_{ij}= {A_j \over r_{ij}}(1+ {A_j \over 24 r_{ij}^2}) \ \ .
\end{equation}
%
The expression for the self-induced potential is 
\begin{equation}
U_{ii}=3.525 \, d           \label{eq:selfUforSquare}
\end{equation}
where $d$ is the element size (i.e. the grid step).

Eq.~(\ref{eq:CMsystemForSigma}) can be  solved numerically and then
the electric field can be computed. However at this point
we face a difficulty. As we have seen earlier, the field
intensity $E$ diverges near the edge of the cluster. In addition,
the charge density itself diverges near the cluster's boundary.
Thus we are trying to find small variations in a diverging quantity,
which itself depends on another diverging quantity.
We can alleviate this problem by introducing line charges
along the boundary. 
That is, in addition to charged square elements, we
will assume that the boundary segments are also charged. This
modification can be easily incorporated into Eq.~(\ref{eq:CMsystemForSigma});
one needs only to distinguish between the self-potential $U_{ii}$
for square elements, which is given by Eq.~(\ref{eq:selfUforSquare}),
and the self-potential $U_{ii}$ for line elements at the cluster's
boundary,
given by 
\begin{equation}
U_{ii} =2 w  \ln (5.44 w/l_i)
    \label{eq:selfUforRod}
\end{equation}
where $w$ is the segment's width (i.e. border width) and
$l_i$ is the length of the segment.
The accuracy can be further improved by using the exact field of a uniformly
charged rod 
rather than the point charge field given by Eq.~(\ref{eq:pointChargeUij}).

As an even further simplification of
the problem, we may use the linear (boundary)
elements {\it only}. The basis
for neglecting the interior elements (which are exactly zero in 3D)
is the fact that the charge density in 2D
is the largest near the boundary, so that charges located
close to the boundary give the main contribution to the electric
field. This simplification not
only reduces the number of elements to consider (and thus
speeds the computation), but also makes it easy to compute
the electric field intensity: to compute a field at distance
$\delta$ from element $i$,
one can just use 
\begin{equation}
  \vec E_i=2\hat n_i \sigma_i w/\delta
\end{equation}
The Boltzmann factor for a neutral particle in this field is then given by
$\exp(\chi E_i^2/(kT))=\exp(\gamma \sigma_i'^2)$,
where $\gamma=4 \chi w^2 \bar \sigma^2/ (\delta^2 k T)$ is the dimensionless
interaction constant, and $\sigma_i'$ is the dimensionless charge density,
determined
by Eq.~(\ref{eq:CMsystemForSigma}) and by the
rescaling condition $\sum (\sigma_i' l_i)/\sum l_i=1$.

We find this simplification to be very important, because, when
the charge density is modeled by a square
grid (even if the grid is very fine), the nonuniformity in the charge density
(arising from the finite grid step) causes
noticeable (artifactual) fluctuations in electric field, which are
then {\it amplified\/} by the growth of the island.
In contrast (as will be seen below), if we locate all the charge on the
boundary elements then the discretization does not lead to 
such artifacts.
Of course, this simplification results in  a less realistic field, but
we believe that the main physical features of the system---the
charge redistribution within the cluster, and the resulting
branch repulsion and competition---are preserved.

We add one further ingredient to our model, namely, the property (i):
branching. The branching instability in growing forms is well known
\cite{7,8,9};
it arises in diffusive problems as a result of competition between 
surface (incoming) diffusion---which tends to favor new branches---and
edge diffusion (along the boundary)---which tends to keep the cluster
smooth. Such competition may be expressed in terms of a length scale
$L_0$; this is essentially the cluster size, or branch length, at which
edge diffusion can no longer maintain the smoothness of the growing
form, and new branches form. Our continuum model does not explicitly 
represent the underlying diffusive processes. Hence we include a
branching instability in a simple way as follows.
When a branch reaches a
size $L_0$ (or after a certain number of growth steps---which 
is approximately the same criterion, for the fastest-growing branches)
we force it to split exactly at the point of the fastest growth
(as determined by the charge distribution) into 2 branches. This branch
splitting is just the simplest implementation of the fingering/tip splitting
instability
which is present in many growth phenomena \cite{fingering0}. This instability
arises near the tips of fast advancing branches; and
the faster the growth, the more likely the split to occur \cite{fingering0}.
For most diffusive problems one may expect branches to form also away from
the fastest-growing tips. We have explored this possibility numerically,
and find in fact that, if we introduce such branches away from the tips,
they do not grow---due to the severe competition present in our model.
Hence, in the following, we only describe cases where noise is
introduced near the tips of the fastest-growing branches.

Obviously, the model described thus far
lacks any source of noise (apart from some
tiny numerical noise, which is always inevitably present).
This means (among other things) that all symmetries present in the
initial shape of the cluster will be preserved during the growth. 
To explore the possibility of chiral symmetry breaking,
then, we need to add some (very small) noise into the system---noise
which breaks 2D inversion symmetry.

Based on the considerations above, our typical
starting condition for numerical studies is an ellipse, with some very
small defects added to represent noise.
As mentioned above, the cases of interest, in which the defects
induce further branching, occur when the noise is added near the
tips---ie, near the long axis of the ellipse.
As one test of the sensitivity of the model to {\it chiral\/} noise,
we have used as a starting form an ellipse
with 2 tiny defects, 
which are small dents placed near (but not at) the tips of the ellipse,
along the long axis. The right dent is placed slightly
above the $x$ axis, and the left dent slightly below.
Thus these defects break 2D inversion symmetry.
Fig.\ \ref{fig:best} shows the results after several generations of
branching. We note that the original
defects are so tiny that they are almost invisible in  Fig.\ \ref{fig:best};
however the original tiny chirality has been increased hugely by the
resulting growth. We also see that Fig.~\ref{fig:best} bears an obvious  
resemblance the experimentally observed ``seahorses''.
In fact, it  has essentially the same geometric properties
as a typical experimental seahorse: 1) only 2 main arms; 2) these main
arms are curved; 3) the curvature is correlated (that is, the two branches
curve in opposite directions); 4) the outer edge of each
main arm is covered by ``fins'' (small dead branches).

It is important to test that such dramatic effects are actually implied
by the dynamics of the model, rather than artifactual
(ie, resulting from some defects in the implementation of the model). 
In fact, it is possible to ``discover'' a chiral symmetry breaking in the 
present model arising simply from the fact that
a certain ordering of points along the boundary (clockwise or
counterclockwise) is present in the numerical algorithm.
Such an ordering can break the left/right
symmetry of the numerical growth results unless sufficient care is exercised. 
We have discovered and removed such artifacts in our own algorithm.
A good test is then a numerical simulation in which 
the starting shape is an ellipse with 2 defects
which {\it do not} break the symmetry (ie, the defects are located 
precisely at the tips of the
ellipse). The result of such a test is  presented in Fig.~\ref{fig:symmetric}:
the original symmetry is retained during subsequent growth.
Hence we are confident that our numerical studies are 
free of any {\it artifactual\/} chiral symmetry breaking.

Now let us consider in more detail how the observed curvature arises.
Fig.~\ref{fig:evolution} gives a more detailed picture of the growth and
branching processes at one end of the island.
After the first split (Fig.~\ref{fig:evolution}a), the new branches compete 
for further growth (as determined by the electric field),
and one of the new branches ``dies'' 
[ie, practically stops growing (Fig.~\ref{fig:evolution}b)]. The winning branch
eventually reaches the critical length and then splits again 
(Fig.~\ref{fig:evolution}c).
Let us consider
what happens when this branch splits in some detail.
First, we note that the new branches ``feel'' one another.
The electrostatic repulsion 
will cause the charges to redistribute, so that the  points
of fastest growth will no longer coincide with tips of the
new branches. Instead, the fastest growth will occur at points of maximal
local charge density (hence maximal $|\vec{E}|^2$), which are
farther away from each other than tips of the branches. This
will force the branches to curve away from one another.
If the split is originally symmetric, then the result of the
competition between the new branches
will be determined by the influence of all the other (already dead) branches.
In particular,  the main body (being closer to the lower branch)  
favors the upper new branch (branch 3), 
while the previous branch (branch 2) favors the lower new branch (branch 4).
Branch 2 is smaller than the main body, but the main body
is further away. Hence, if the parameter values are right (in particular, if
the distance between splitting points is small enough, and dead branches
are large enough), then the nearby dead branch has a dominant
influence, and so
branch 4  will win the competition. Then, due to the exponential difference
in the growth rate,  branch 2 will stop growing. If this sequence is
then repeated, the main arm curves further to the right at each branching.

A similar process happens on the opposite end of the cluster.
Hence we see how the present model leads to two main arms, each of which
may be rather strongly curved. An `S' shape (rather than a `C'
or a `3' shape) results if the direction of curvature is the same
for both main arms; that is, the two arms must show a {\it
correlation\/} in curvature. This implies that one end of the growing
body is significantly affected by the form of the distant end. 
We believe that this point is important: curvature itself, such as shown
in Fig.\ \ref{fig:evolution}, may be ascribed to relatively {\it local\/}
effects. However the correlation giving rise to consistent `S' shapes 
appears to {\it require\/} some form of long-ranged communication 
between the parts of the growing island. It is interesting to note here
that, in a small minority of the experimentally observed islands,
there are two arms whose curvature is in opposing directions---giving a
`3' form rather than an `S' \cite{gao3}.

This correlation is not tested by our Fig.\ \ref{fig:best}, since it is
built into the starting condition. We demonstrate that such a
correlation does occur in the present model in Fig.\
\ref{fig:correlation}. Here the starting condition is an ellipse with
one (right) defect off axis (to break the symmetry), but the other one
precisely on axis.
We see that the growth dynamics not only causes the right arm of the cluster
to curve downwards, but also determines the outcome of the competition
at the other end of the island. It is clear from the figure that the
growth at this end is initially symmetric; yet subsequently, the upper
arm dominates, purely due to influence from the other end of the cluster.
We see that `clockwise' growth at one
end has sufficient influence to force clockwise growth at the other.
Thus we find that dominance and curvature {\it are\/} correlated
between the two main arms,  leading to strong chirality of the 
overall cluster.

We have also examined the case that both defects are on the same side of
the main axis. Here too we find that the two branches, while initially
curving as expected from the local bias, after some growth `feel' one
another and so are repelled. We believe however that this starting condition 
is less realistic than those in Figs.\ \ref{fig:best} and
\ref{fig:correlation}, because the noise we introduce into our
continuum model does not represent truly microscopic noise, but rather
that noise that may be expected to grow beyond the microscopic scale.
This is because the continuum model itself is not a truly microscopic
model. Below (section \ref{sec:MC}) we will give evidence
from our microscopic Monte Carlo
studies that this continuum model does indeed capture 
important features of the
large-scale behavior of a microscopic model with noise---in particular,
the two-arm instability, and the repulsion between these two arms.

It is entirely plausible that a ``non-local'' effect such as 
the correlation of curvature shown in Fig.~\ref{fig:correlation} is
more fragile than the more local effects shown in Fig.\ \ref{fig:evolution}.
This is true experimentally \cite{gao3}: 
as noted above, while the large majority of individual
experimental clusters form an `S' shape, some do not.
Yet it is precisely this kind of long-distance correlation that gives
rise to the broken inversion symmetry: in the absence of such
correlation, half of the two-armed islands would have a `3'
or `C' shape---and these shapes are not 2D chiral.

Finally, we wish to demonstrate in yet another way that the dramatic 
symmetry breaking shown in Fig.\ \ref{fig:best} is a genuine outcome of
the growth dynamics of our model:
that is, the `S'
shape is not a {\it necessary\/} outcome of the growth model, yet
neither does it require fine tuning of the model parameters. 
Starting with the same initial form as that leading to Fig.\
\ref{fig:best}, 
we have performed simulations for different values of the two
model parameters: $L_0$
(the branching length), and the electrostatic
interaction constant $\gamma$. The results
are presented in Fig.~\ref{fig:wiremap}.

A number of conclusions may be drawn from Fig.~\ref{fig:wiremap}.
We see clearly that {\it branching\/} is important for obtaining S
shapes: in the bottom row of the figure, branching is essentially absent,
and the resulting forms amplify the broken symmetry of the initial
condition only rather weakly. We also see that charge effects are
equally important. They are weak in the left-hand column of 
Fig.~\ref{fig:wiremap}; and the resulting forms are uninteresting,
regardless of the branching length. Finally, we note that, while both of
these effects are important---in that they must be present in order that
`interesting' growth forms result---fine tuning of the two model 
parameters is {\it not\/} needed: we see clear two-armed `S' shapes over
a central region of the parameter space.

In short: Fig.~\ref{fig:wiremap}---along with the previous
figures---demonstrates rather clearly that 
growth dominated by (electrostatic effects $+$ branching) {\it can\/}
give rise to spontaneous `S' shapes which strongly break 2D inversion
symmetry, and that these forms are {\it not\/} 
simply modeling artifacts arising either from fine tuning or from
being built into the model by hand.

We thus assert that our model possesses a genuine
{\it chiral} growth instability; that is, it can drastically amplify 
any tiny perturbation that breaks the left/right symmetry. 
Such `drastic amplification' is of course what is meant by the term
`spontaneous symmetry breaking'---which is also an appropriate
description. Hence the kind of pattern-forming behavior studied 
here falls into the same, quite rare class of behavior as that seen in
phyllotaxis \cite{phyllotaxis,Douady}.

\section{Monte Carlo studies}
\label{sec:MC}

We have performed a number of MC studies of aggregation under
conditions where the aggregate is charged, and there are both charged
and neutral walkers. The most obvious obstacle to overcome is the
size of the clusters; as mentioned earlier, there are $10^6$--$10^7$ 
particles in a typical experimental cluster.
Thus, in order to perform a successful MC simulation of these
experiments, one must find a regime where the desired behavior
(formation of S-like clusters) is achieved with smaller numbers of particles.

We have explored a number of MC models, involving various combinations
among the following choices: charge-conserving or not (as above); 
following each walker individually, or making (again) a quasiequilibrium
assumption for the spatial distribution of walkers; 
and treating the cluster as insulating or conducting.
Our most encouraging results were obtained using the latter choice
in each of the above three cases: that is, modeling only the neutral
walkers, treated as a quasiequilibrium gas bonding to a conducting
cluster. The resulting model is very like our continuum model in many
ways---with the important difference that branching is now a part of the
model dynamics, rather than being enforced as a premise of the model.

With the above assumptions, the growth rule is as follows. 
We perform the growth simulation with the underlying 2D space
discretized as a triangular lattice.
Given a shape of the cluster (ie a set of occupied sites---our starting
cluster is typically a single site),
we first find all non-occupied lattice sites adjacent to the cluster. Then, for
all sites simultaneously, we compute the non-normalized probability of a 
new particle joining the cluster at this site according to
\begin{equation}
   P_s=A\, \exp[(\chi E^2_s/2 + K_s E_b)/kT]  \label{eq:semilocalRule}
\end{equation}
where subscript $s$ numbers sites along the cluster boundary, $E_s$ is
the 
electric field strength at site $s$, $A$ is
a normalization constant (effectively, the time scaling constant), $K_s$ is
the number of sites neighboring site $s$ which 
are already occupied, and $E_b$ is the  bond
energy. Thus, again we have a two-parameter model. The strength of 
electrostatic effects is set by the dimensionless $\chi$. The
competition of surface and edge diffusion is contained in the bond
energy $E_b$: large $E_b$ tends to keep the cluster smooth, while
small $E_b$ allows small irregularities to grow.

Using this MC model, we were able to reproduce some of the features of
the seahorse clusters. Specifically, we found a clear two-arm
instability for a significant area of the two-dimensional parameter
space. We also found some tendency for the arms to curve, with 
small dead branches on the outer edges of the curves
(Fig.~\ref{fig:mcbest}). However---recognizing 
now that the particle number $N$ represents in a sense a third dimension
to be explored (with our present range of $N\alt 2000$)---we did not 
find a volume of this 
three-dimensional space in which the curvature appeared consistently.
It is however interesting to note that, where curvature did appear,
there was an apparent tendency for the curvature of the two arms to be
correlated.

As mentioned above, in tests of our continuum model we found that there
was a four-arm instability, which however was weaker than the two-arm
instability over a range of growth parameters. One can also see this
from Fig.\ \ref{fig:wiremap}: for some parameters, one finds four
dominant arms, and for others, two. Here we see the same 
competition in Fig.\ \ref{fig:mcbest}: two arms have grown to dominance,
yet there are two others which have not entirely stopped growing.
We have also plotted out a two-dimensional `map' of MC growth patterns,
depending on the electrostatic parameter $\chi$ and the smoothness
parameter $E_b/kT$. We find that this `map' is qualitatively like that
shown in Fig.\ \ref{fig:wiremap}: compact shapes for small $\chi$
and large $E_b/kT$ (roughly equivalent to small $\gamma$ and large $L_0$
in the continuum model);
and two, relatively straight, arms for large $\chi$ and small $E_b/kT$.
There are also some significant differences between the two maps.
One difference is (again) that we find no region [in $(\chi,E_b)$ space,
at $N\sim 2000$] of consistent
curvature of the two arms, giving S shapes. Another is some tendency for
three or six arms to appear; this is almost certainly an artifact of the
underlying triangular lattice.

Clearly these results support some important aspects of the picture
gained from our other, less microscopic models, even as they fail to
give `seahorse' forms.
The two-arm instability, with the tendency for the two
arms to lie $\approx 180^\circ$ apart, is clear; also the microscopic parameters
$\chi$ and $E_b/kT$ give qualitatively the same growth behavior as do 
$\gamma$ and $L_0$ in the continuum model. However we have not found
consistent S shapes from the MC models that we have explored.
We believe that MC simulations with larger $N$
are needed to test these ideas and results further.
It is clear that the kind of repeated branching and death shown in
Fig.\ \ref{fig:evolution} requires a minimal number of microscopic
particles in an aggregate before it can appear. Each dead branch 
which plays an important role in driving the curvature
must be formed of some minimum number of particles; and there must
be several generations of dead branches---and of course concomitant growth
of the entire aggregate---for the curvature to become
significant. We believe that
our current results are below that threshold in $N$
(assuming that it exists). Our current MC results also
show too much dominance from both noise and the underlying lattice.
It is clear from Fig.\ \ref{fig:mcbest} that artifacts from the
underlying lattice are not negligible at the small scale of the figure.
Also, microscopic noise is still large at this scale, 
compared to the other physical effects influencing the growth.
Both of these effects will become less important at larger $N$.

\section{Growth in an external electrostatic field}

The results that we have presented above represent an unusual
form of symmetry breaking in the modeling of growth phenomena.
While these studies were inspired by the same unusual symmetry
breaking seen in the ICB experiments \cite{gao1,gao2,gao3}, it is by no means
certain that the mechanisms explored here are in fact responsible for
the observed growth patterns. Our ideas have however motivated further
growth experiments, to be reported in detail in a separate publication
\cite{gao3}. Here we will briefly describe the idea of the experiments,
and the corresponding simulation studies that we have done, using our
continuum model.

ICB growth experiments have been performed as in Refs.\ \cite{gao1}
and \cite{gao2}, with the single change that there is imposed an
external electrostatic field in the plane of the films. The motivation
is to test whether electrostatic effects are indeed important for the
growth forms; and the results \cite{gao3} say that they are.
Qualitatively, the in-plane field tends to give three effects:
(a) the curvature of the main arms is reduced by the field, and even
eliminated in a sufficiently strong field; (b) growth appears to be
predominantly at one of the two main arms; (c) there is a {\it weak\/}
tendency (which cannot be distinguished with certainty from zero)
for the straightened clusters to grow in alignment with the external
field.

It is straightforward to include an external electric field in our
continuum simulations, by adding another term to the equations
(\ref{eq:CMsystemForSigma}) for the charge densities.
A typical result from such a simulation 
is presented in Fig.~\ref{fig:external}.
The starting condition is the same as for Fig.~\ref{fig:best}.
We see that the cluster has lost most of its curvature. Also
there is
some competition between the tendency to curve (which is favored by the
starting defects) and the external field (which tends to align
the growth along the field direction). The result of this competition
is (roughly) a straight branch, growing at an angle to the field. 
These results show a good qualitative agreement with
the experiments: (a) the curvature of the main arms is suppressed; 
(b) only one arm grows; and (c) there is some weak tendency for growth
to follow the direction of external field, and at the same time, some
reason to expect the growth to deviate from the orientation of the
external field.

\section{Conclusions}

Despite the obvious resemblance between the experimental results and
our simulations, many questions need to be answered before one can claim
that our model is directly related to the experiment. 
First, TCNQ crystals are not
conducting \cite{TCNQconduction1,TCNQconduction2}. However they are believed to 
have a large (about $8$) dielectric constant 
\cite{TCNQepsilon}. 
Obviously, the polarization charge distribution on the surface
of a dielectric will be different from the charge distribution on 
the surface of a conductor; but 
the larger the dielectric constant, the smaller this difference is.
Second, our model relies crucially on the  assumption of slow growth (this
assumption leads to exponential differences in growth rates). It is not clear
how close  this assumption is to what really happens in the experiments.
Third, there are very large uncertainties regarding the likely
behavior of $Q(t)$, the charge on an island as a function of time.
Finally, there is one feature in the experimental patterns that can
almost certainly not be obtained from our electrostatic mechanism alone,
namely, in the experimental seahorses the curvature is strong enough and
prolonged enough that the main arms commonly bend back to touch the
central body of the island. We speculate that the addition of crystalline
anisotropy effects at grain boundaries can yield this kind of behavior.

One might argue that some other kind of long-range force---for example
that coming from elastic effects---might produce an S shape. Of course
we cannot rule out this possibility. However the physical effects
expected from elastic forces are rather different from those explored
here, arising from electrostatic effects. Elastic stresses are not
concentrated at growing tips, but rather at `valleys' of the solid's
boundary. Also, in our picture the electrostatic effects are important
beyond the boundary of the aggregate. Elastic forces can also extend
through the substrate, beyond the boundary of a growing island, during
epitaxial growth on a crystalline substrate \cite{elastic}; however
such effects seem likely to be much smaller for an amorphous substrate
such as that used in the ICB deposition experiments.
Finally, it is not clear to
us how elastic forces could give the broken symmetry seen
experimentally, nor the strong sensitivity of the growth patterns to an
external electric field \cite{gao3}.
We do note however that some {\it combination\/} of elastic and electrostatic
effects may be needed to give rise to the high degree of `bending back'
of the main arms (noted above) seen in the experimental seahorses.

We comment upon our modeling results.
It remains to be demonstrated that all of the features
reported here (two arms, branching, consistent correlated
``deaths" of branches leading to curvature, and correlation between the
curvatures of the two main arms) can be seen to occur together
in a single model, with none of them being imposed on the model
by construction. Our continuum model shows all of these features but 
the branching itself (which is uncontroversial); 
and our MC studies have shown hints (or more) of all of these
features. We believe that all of our results, considered
together, suggest rather strongly that the principal physical idea
explored in our various models---trapped electric charge on a growing
aggregate, leading to strongly enhanced growth in regions of strong
electric field---can indeed lead to all of these features. However 
the final demonstration of this remains a challenge to future 
work in theory and modeling.

In summary: we have proposed a novel mechanism for spontaneous chiral 
symmetry breaking during 2D aggregation. This mechanism does not rely 
on the presence of microscopic chirality; instead, it results from 
the existence of a long-ranged electrostatic
interaction in the system, due to a trapped charge on the growing
aggregate. This interaction leads to a strong competition
and repulsion between growing branches, and---as we have shown
here---can give rise to a strongly but {\it spontaneously\/} broken
2D inversion symmetry. We have explored several different
approaches to the simulation of growth in the presence of such 
an electrostatic interaction: tree models, continuum models,
and Monte-Carlo simulations. Especially encouraging results were obtained using
the continuum model; simulated growth often led to clusters which possess
the same geometrical properties as the experimentally observed ``seahorses''
\cite{gao1,gao2}. The results of our simulations in the presence of an
external electric field are also in good qualitative
agreement with the experimental
results. Further theoretical work, and further experiments, are 
needed in order to clarify and test
the connection between the ideas presented here
and the ``seahorse'' experiments. However, using physical ideas motivated
by the growth experiments, we have found a new class of `electrostatic
growth' models which spontaneously break 2D inversion symmetry during growth;
and we remain convinced that such spontaneous symmetry breaking occurs in the 
``seahorse'' experiments, and so demands a theoretical model 
which does the same.

{\it Acknowledgments.}--
GSC thanks Andreas Deutsch for
helpful discussions of bacterial aggregation.
This work was supported in
part by the U.S. Department of Energy through Contract No.
DE-AC05-96OR22464 with Lockheed Martin Energy Research Corp.
IMS and GSC were supported by the NSF under Grant \# DMR-9413057.


\begin{thebibliography}{10}
\bibitem{uncommon} For example, Ref.~\cite{fractalGrowth} describes
over 150 two-dimensional growth forms, yet only one (bacterial) growth form 
is 2D chiral (Plate 13); and here the chirality was forced by an experimental
constraint.
\end{thebibliography}

\begin{figure}
\caption{The first two steps of a charged tree's growth. 
At every step, the tree selects
a configuration with the lowest electrostatic energy. For the first step 
one of the rejected configurations is shown above the transition arrow.}
\label{fig:treegrowth}
\end{figure}

\begin{figure}
\caption{The growth of a conducting tree. $l_1=0.7$, $l_0=0.5$ (arbitrary
length units); $\theta=12^\circ$. In this case, growth according to an
energy-minimization rule leads to an `S' shape.}
\label{fig:besttree}
%
\end{figure}

\begin{figure}
\caption{Sixty time steps in the growth of a charged island.
The starting condition is an ellipse with two tiny defects, each placed
just counterclockwise of the ellipse's symmetry axis. The tiny symmetry
breaking present in these defects is
dramatically enhanced by the subsequent growth.}
\label{fig:best}
\end{figure}

\begin{figure}
\caption{Growth of an ellipse with 2 tiny {\it symmetrically} located defects.
Such a  shape preserves its symmetry during the growth. Hence the
symmetry breaking seen in the other figures is not artifactual.}
\label{fig:symmetric}
\end{figure}

\begin{figure}
\caption{Early stages in the growth of the right half of Fig.~3,
viewed in detail. (a),(b) The broken symmetry in the original (tiny) defect
leads to dominance of the lower branch. (c) The growing branch 1 splits
again. (d) The nearby ``dead'' branch 2 inhibits branch 3, so that
again the lower branch dominates. In this way the `memory' of the
{\it handedness\/} (right or left) of the 
original small defect is maintained and amplified by the subsequent
growth.}
\label{fig:evolution}
\end{figure}

\begin{figure}
\caption{Growth from an initial shape in which one defect (on the
right) is off the symmetry axis, but the other (on the left) is on the
symmetry axis. The broken symmetry on the right-hand side still has a
large effect on the growth of the left-hand side---in fact, it causes
the same (clockwise) branch to dominate there, even though the
left-side defect is symmetrically placed.}
\label{fig:correlation}
\end{figure}

\begin{figure}
\caption{Growth forms for different values of the branching length $L_0$ 
(increasing downwards) and the interaction constant $\gamma$
(increasing to the right). There is a region of the figure
(roughly, the center) in which growth gives `S' forms resembling those
seen experimentally. However if the electrostatic effects are too weak
(left side) or branching is too infrequent (bottom) then other forms
result, displaying little or no chiral symmetry breaking.}
\label{fig:wiremap}
\end{figure}

\begin{figure}
\caption{A 2300-particle cluster grown using a Monte Carlo
algorithm on a triangular lattice. The growth rule is described in the
text; growth parameters are
$\chi=1.0$ and $E_b/kT=2.5$. Although this cluster looks promising,
we have not obtained such results consistently. It is probably
necessary to use larger particle numbers to obtain such forms
consistently.}
\label{fig:mcbest}
\end{figure}

\begin{figure}
\caption{Growth in the presence of an external electrostatic field, 
oriented horizontally to the right. Starting conditions and growth rule are
otherwise as in Fig.~3.}
\label{fig:external}
\end{figure}

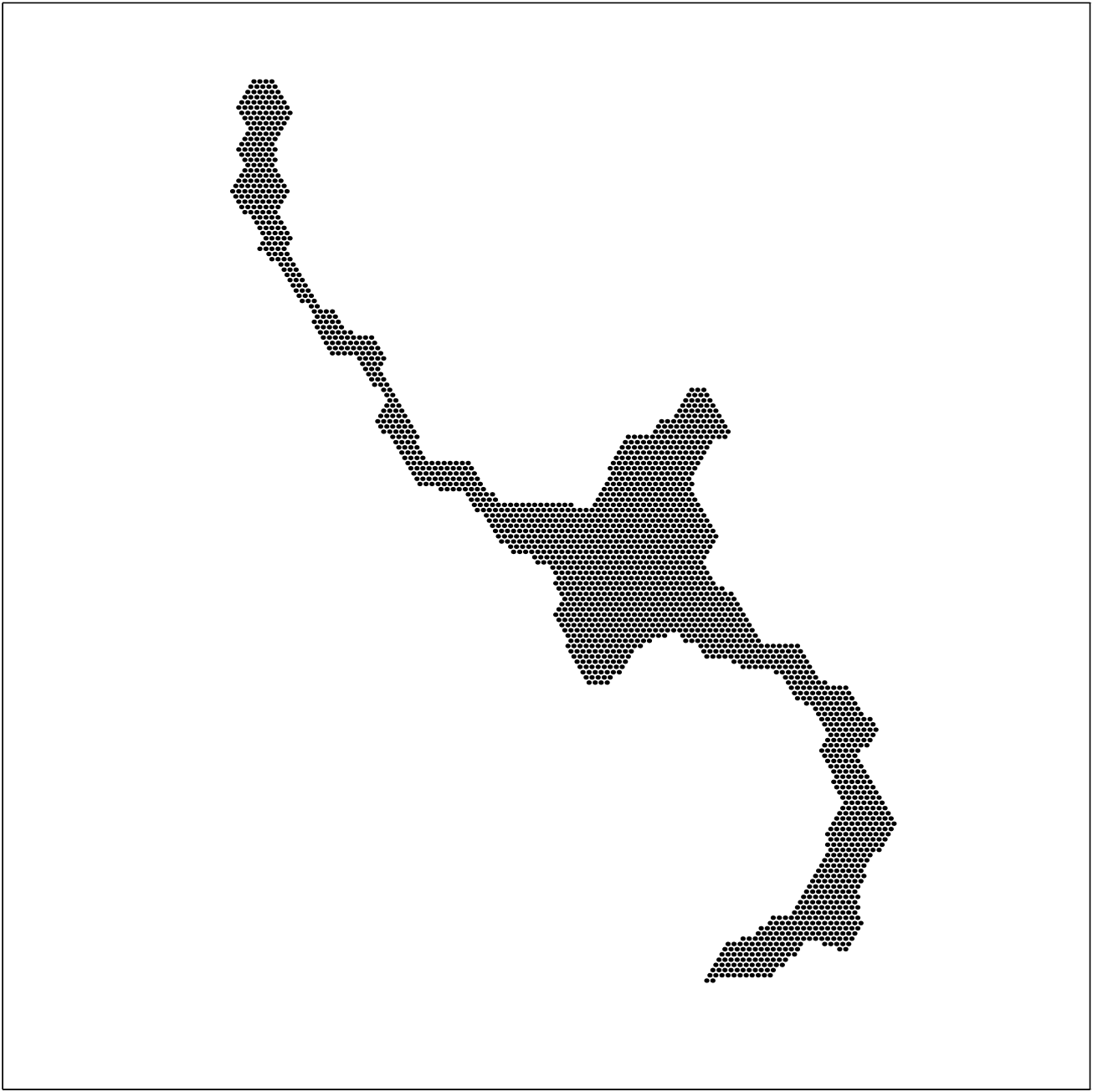
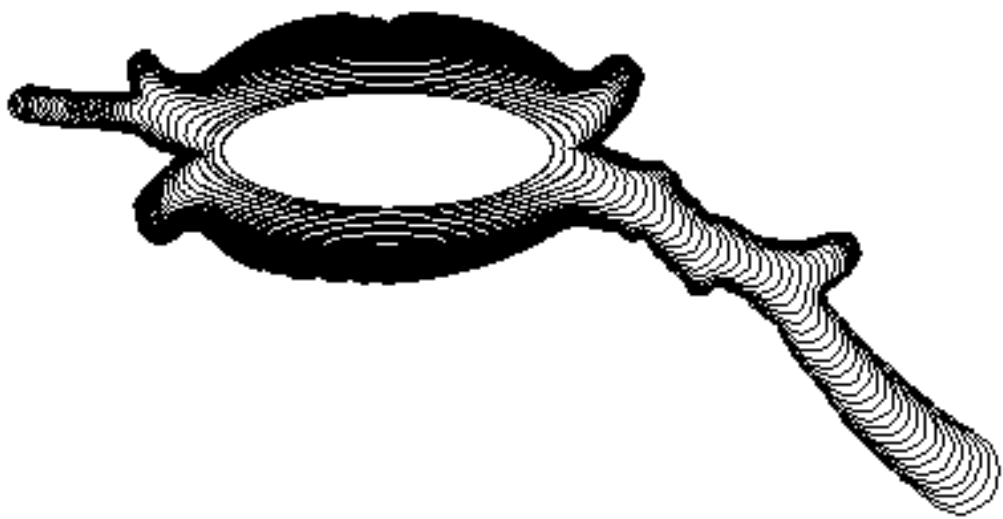
\end{document}